\documentclass[pra,twocolumn,showpacs,floatfix]{revtex4}
\usepackage{graphicx}

\newcommand{\figwidth}{0.95\columnwidth}

\newcommand{\eq}[1]{Eq.(\ref{#1})}
\newcommand{\fig}[1]{Fig.~\ref{#1}}
\newcommand{\tab}[1]{Table~\ref{#1}}
\newcommand{\olcite}[1]{Ref.~\onlinecite{#1}}
\newcommand{\avg}[1]{ {\langle #1 \rangle} }

\newcommand{\Rc}{ R_{\rm c} }
\newcommand{\Rp}{ R_{\rm p} }
\newcommand{\Nc}{ N_{\rm c} }
\newcommand{\Np}{ N_{\rm p} }
\newcommand{\zc}{ z_{\rm c} }
\newcommand{\zp}{ z_{\rm p} }
\newcommand{\etac}{ \eta_{\rm c} }
\newcommand{\etap}{ \eta_{\rm p} }
\newcommand{\etapr}{ \eta_{\rm p}^{\rm r} }
\newcommand{\etaccr}{ \eta_{\rm c,cr} }
\newcommand{\etapcr}{ \eta_{\rm p,cr} }
\newcommand{\etaprcr}{ \eta_{\rm p,cr}^{\rm r} }
\newcommand{\Pc}{ P(\etac) }
\newcommand{\etacv}{ \etac^{\rm v} }
\newcommand{\etacl}{ \etac^{\rm l} }
\newcommand{\wv}{ W_{\rm v} }
\newcommand{\wl}{ W_{\rm l} }

\begin{document}

\title{Critical phenomena in colloid--polymer mixtures: interfacial 
tension, order parameter, susceptibility and coexistence diameter}

\author{R. L. C. Vink, J. Horbach, and K. Binder}

\affiliation{Institut f\"{u}r Physik, Johannes Gutenberg-Universit\"{a}t,
D-55099 Mainz, Staudinger Weg 7, Germany}

\date{\today}

\begin{abstract}

The critical behavior of a model colloid--polymer mixture, the so--called
AO model, is studied using computer simulations and finite size scaling
techniques. Investigated are the interfacial tension, the order parameter,
the susceptibility and the coexistence diameter. Our results clearly show
that the interfacial tension vanishes at the critical point with exponent
$2\nu \approx 1.26$. This is in good agreement with the 3D Ising exponent.  
Also calculated are critical amplitude ratios, which are shown to be
compatible with the corresponding 3D Ising values. We additionally
identify a number of subtleties that are encountered when finite size
scaling is applied to the AO model. In particular, we find that the finite
size extrapolation of the interfacial tension is most consistent when
logarithmic size dependences are ignored. This finding is in agreement
with the work of Berg {\it et al.} [Phys. Rev. B, {\bf 47}, 497 (1993)].

\end{abstract}



\maketitle

\section{Introduction}

By adding non-adsorbing polymer to a colloidal suspension, phase
separation may be induced. This leads to the formation of two coexisting
fluid phases, separated by an interface~\cite{gast1983a, poon1998a}. The
phases are characterized by their colloid density, which is high in one
phase, and low in the other. In order to make the analogy to the
fluid--vapor transition in atomic liquids, the colloid rich phase is
usually called the colloidal liquid, and the colloid poor phase the
colloidal vapor. Obviously, the density of the polymers is exactly the
opposite: high in the colloidal vapor and low in the colloidal liquid.

In the vicinity of the critical point, a number of important physical
quantities are described by simple power laws of the form $A t^B$. Here,
$t$ is some measure of distance from the critical point, $A$ is called the
critical amplitude, and $B$ the critical exponent~\cite{binder2001a}. To
describe the critical phase behavior, one thus needs to determine the
location of the critical point, the critical exponents and the critical
amplitudes. These quantities can for instance be obtained in computer
simulations, provided finite size scaling methods are used. Finite size
scaling is required because the correlation length diverges at the
critical point. Since the accessible system size in a simulation is
finite, the true critical behavior is obscured as soon as the correlation
length exceeds the size of the simulation box~\cite{deutsch1992a,
binder2001a}. Finite size scaling offers a way to properly extrapolate the
data obtained in simulations to the thermodynamic (=infinite system)
limit~\cite{lb_book}.

In this work, we study the critical behavior of the colloid--polymer model
introduced by Asakura and Oosawa~\cite{asakura1954a, vrij1976a} (the
so--called AO model). In previous simulations, we have determined the
critical point of the AO model (for one choice of colloid--to--polymer
size ratio) and we have provided evidence that this system belongs to the
3D Ising universality class~\cite{vink2004a, vink2004b}. However, we have
not yet studied in detail whether we can recover the critical exponents,
nor have we determined the critical amplitudes. The latter quantities are
of interest because certain critical amplitude ratios are predicted to be
universal. In this work we combine computer simulations and finite size
scaling to address these issues. The quantities that we consider are the
order parameter, susceptibility, coexistence diameter, and interfacial
tension, whereby the critical point is approached along different paths:
from the one--phase region, the two--phase region, and along the
coexistence line.

The AO interfacial tension is of particular importance, because it gives
an indication of the strength of capillary waves. This is an issue in
(mean--field) density functional theories of the AO model. Following
\olcite{brader2003}, the strength of the capillary waves is estimated by
$\omega = k_{\rm B} T / (4 \pi \sigma \xi^2)$, with $\sigma$ the
interfacial tension, $\xi$ the correlation length in the two--phase
region, $T$ the temperature, and $k_{\rm B}$ the Boltzmann constant. For
3D Ising critical behavior, hyperscaling implies that $\omega$ is constant
in the critical regime. Note that this constant is universal and given by
$\omega \approx 0.8$~\cite{pelissetto2002}. In contrast, for mean--field
critical behavior, one would observe a decay of the form $\omega \propto
t^{-1/2}$. The mean--field behavior and the 3D Ising behavior of the
capillary strength are thus profoundly different. Therefore, it is
important to establish the universality class of the AO model. For this
purpose, an analysis of the interfacial tension is particularly suitable.
To determine the critical behavior of the interfacial tension, finite size
scaling methods can be used that do not require prior knowledge of the
universality class. This enables a direct measurement of the critical
exponent, and the corresponding critical amplitude. For the AO model, we
obtain for the interfacial tension a critical exponent $2\nu=1.26$, which
is in excellent agreement with the 3D Ising value.

The critical amplitudes are used to test the universality of a number of
critical amplitude ratios. Following Stauffer~\cite{stauffer1972,
privman1991a}, universality implies that only two amplitudes are required
to determine the remaining amplitudes. This allows us to relate the
critical amplitudes of the AO model obtained in this work, to independent
estimates obtained in experimental, theoretical, and simulational studies
of completely different systems~\cite{privman1991a, pelissetto2002}. We
observe reasonable agreement, but emphasize that the error bars, in both
our estimates and those in the literature, are quite substantial.

Note that an investigation of the AO critical behavior is far more complex
than an equivalent study of the 3D Ising lattice model would be. For
instance, the binodal of the AO model is asymmetric, and this gives rise
to an additional critical power law for the coexistence diameter.
Moreover, the AO model belongs to the class of {\it asymmetric binary
mixtures}, which are generally difficult to simulate. In case of the AO
model, the accuracy required to apply finite size scaling became available
only after the recent introduction of a grand canonical Monte Carlo
cluster move~\cite{vink2004b} and successive umbrella
sampling~\cite{virnau2003a}. Note also that the application of finite size
scaling to asymmetric mixtures~\cite{muller1995a} is far less common
compared to that of symmetric mixtures~\cite{deutsch1992a}.

The outline of this paper is as follows. First, the AO model is
introduced. We then move on to describe the simulation techniques used by
us. Next, we explain how the order parameter, susceptibility, interfacial
tension and coexistence diameter are extracted from the simulation data.
The extrapolation of these quantities to the infinite system is discussed
in section~\ref{sec:fss}. We then present our results and end with a
summary in the last section.

\section{The AO model}

The AO model was proposed in 1954~\cite{asakura1954a} and later independently by
Vrij~\cite{vrij1976a} as a simple description for colloid--polymer mixtures. In
this model, colloids and polymers are treated as spheres with respective radii
$\Rc$ and $\Rp$. Hard sphere interactions are assumed between colloid--colloid
(cc) and colloid--polymer (cp) pairs, while polymer--polymer (pp) pairs can
interpenetrate freely. This yields the following pair potentials:
\begin{eqnarray}
\label{eq:ao}
u_{\rm cc}(r) &=& \left\{
    \begin{array}{ll}
    \infty & \mbox{for $r<2\Rc$} \\
    0      & \mbox{otherwise,}
    \end{array}
  \right. \quad \nonumber \\
u_{\rm cp}(r) & = &
  \left\{
    \begin{array}{ll}
    \infty & \mbox{for $r<\Rc+\Rp$} \\
    0      & \mbox{otherwise,}
    \end{array}
  \right. \\
u_{\rm pp}(r) &=& 0 \nonumber,
\end{eqnarray}
with $r$ the distance between two particles. 

Since all allowed AO configurations have zero potential energy,
temperature plays a trivial role, and the phase behavior is set by the
colloid to polymer size ratio $q \equiv \Rp/\Rc$ and the fugacities
$\{\zc,\zp\}$ of colloids and polymers, respectively. The fugacity
$z_\alpha$ is related to the chemical potential $\mu_\alpha$ via $z_\alpha
= \exp( \beta \mu_\alpha)$, with $\alpha \in \{ {\rm c,p} \}$. 

In this work we consider a size ratio $q=0.8$ and put $\Rc \equiv 1$ to
set the length scale. The colloid packing fraction is defined by $\etac
\equiv (4\pi/3)\Rc^3 \Nc/V$, and the polymer packing fraction by $\etap
\equiv (4\pi/3)\Rp^3 \Np/V$. Here, $\Nc$ ($\Np$) denotes the number of
colloids (polymers) inside the simulation cell and $V$ the volume of the
simulation cell. Following convention, we use the quantity $\etapr \equiv
\zp (4\pi/3)\Rp^3$ to express the polymer fugacity, rather than $\zp$
itself. In the literature, $\etapr$ is known as the polymer reservoir
packing fraction. It should, however, not be confused with the actual
polymer packing fraction in the system $\etap$.

The AO model phase separates into a colloidal vapor and colloidal liquid,
provided $q$ and $\etapr$ are high enough~\cite{lekkerkerker1992a, meijer1994a,
schmidt2000a, bolhuis2002a, schmidt2003a}. For $q=0.8$, the critical point was
located at~\cite{vink2004a,vink2004b}:
\begin{eqnarray}
\label{eq:old}
  \etaprcr = 0.766 \pm 0.002, \hspace{5mm} 
  \etapcr = 0.3562 \pm 0.0006, \\
  \etaccr = 0.1340 \pm 0.0006, \hspace{5mm} 
  \mu_{\rm c,cr}=3.063 \pm 0.003, \nonumber
\end{eqnarray}
with $\mu_{\rm c,cr}$ the critical value of the coexistence chemical
potential of the colloids and $\eta_{\alpha,{\rm cr}}$ the critical value
of $\eta_\alpha$, with $\alpha \in \{ {\rm ^r_p,p,c} \}$. The above
estimates were obtained using the cumulant intersection
method~\cite{binder1981a}, and by considering field mixing
effects~\cite{goldstein1985a, wilding1996a}.

\section{Simulation Method}

We simulate the AO model in the grand canonical ensemble. In this ensemble, the
fugacities $\{z_c,\zp\}$ and the volume $V$ are fixed, while the number of
particles inside $V$ fluctuates. The simulations are performed in cubic boxes
with edge length $L$ and using periodic boundary conditions. To simulate the AO
model efficiently, we use a recently developed cluster move~\cite{vink2004b,
vink2004c}. 

\subsection{Phase coexistence}

During the simulation, we measure the probability $\Pc$ of observing a certain
colloid packing fraction $\etac$.  At phase coexistence, the distribution $\Pc$
becomes bimodal, with two peaks of equal area for the colloidal vapor and liquid
phase. A natural cut-off to separate the vapor from the liquid phase is provided
by the average colloid packing fraction:
\begin{equation}
\label{eq:avg}
 \avg{\etac} = \int_0^\infty \etac \Pc {\rm d}\etac,
\end{equation}
where we assume $\Pc$ has been normalized to unity:
\begin{equation}
\label{eq:norm}
  \int_0^\infty \Pc {\rm d}\etac = 1.
\end{equation}
The equal area rule simply implies that:
\begin{equation}
\label{eq:area}
  \int_0^\avg{\etac} \Pc {\rm d}\etac = 
  \int_\avg{\etac}^\infty  \Pc {\rm d}\etac.
\end{equation}
The above equation provides an accurate numerical measure to determine phase
coexistence~\cite{muller1995a}. 

\subsection{Successive umbrella sampling}

Close to the critical point, the simulation moves back and forth easily
between the vapor and liquid phases. Away from the critical point, at
higher polymer fugacity, the free energy barrier between the two phases
increases. In that case, transitions from one phase to the other phase
become more and more unlikely, and the simulation will spend most time in
only one of the two phases. A crucial ingredient in our simulation is
therefore the use of a biased sampling technique called successive
umbrella sampling. This technique was recently developed by Virnau and
M\"uller~\cite{virnau2003a} and its purpose is to enable sampling in
regions where $\Pc$, due to the free energy barrier separating the phases,
is very low.

\subsection{Histogram reweighting}

A final ingredient in our simulation is the use of histogram
reweighting~\cite{ferrenberg1988a}. It is based on the observation that the
probability $\Pc$ measured at one set of model parameters (in this case $\zc$
and $\zp$) can be used to estimate $\Pc$ at different values of these
parameters. Obviously, the gain in computational efficiency is enormous because
in the ideal case $\Pc$ need only be measured once. 

In this work, histogram reweighting is used to locate the coexistence fugacity
of the colloids. Phase coexistence is only obtained if the colloid fugacity is
chosen just right. This value is in general not known at the start of the
simulation. However, once $P(\etac|\zc,\zp)$ has been measured at colloid
fugacity $\zc$ and polymer fugacity $\zp$, histogram reweighting can be used to
obtain $P(\etac|\zc',\zp)$ at any other colloid fugacity $\zc'$ by using the
equation~\cite{lb_book}: 
\begin{equation}
\label{eq:ext}
  \ln P(\Nc|\zc',\zp) = \ln P(\Nc|\zc,\zp) + 
  \left( \ln \frac{\zc'}{\zc} \right) \Nc.
\end{equation}
Note that in the above equation $\etac$ has been replaced by the number of
colloids $\Nc$. In the simulations, we thus set the colloid fugacity to unity
and apply successive umbrella sampling to obtain the corresponding probability
distribution. We then use \eq{eq:ext} to extrapolate $\Pc$ to that colloid
fugacity at which the equal area rule of \eq{eq:area} is obeyed. In practice, it
is straightforward to write an automated numerical procedure to achieve 
this. We emphasize here that extrapolations in $\zc$ are exact, in the sense
that no statistical or systematic errors are introduced (the term
``extrapolation'' might suggest otherwise).  

Within successive umbrella sampling, states (or windows) are sampled one
after the other. In the first window, the number of colloids $\Nc$ is
allowed to fluctuate between 0 and 1, in the second window, $\Nc$ is
allowed to fluctuate between 1 and 2, and so on. No restriction is put on
the number of polymers though, so in each window $\Np$ will fluctuate
freely around some average equilibrium value (the distribution in $\Np$ is
to a good approximation Poisson like). The crucial point is that by using
successive umbrella sampling, data over the entire range from $\Nc=0$ up
to some maximum is obtained (the maximum should be chosen well beyond the
liquid peak). Therefore, extrapolations in $\zc$ (which essentially
``emphasize'' the data of some windows with respect to others) can be
performed without loss of accuracy.

Histogram reweighting is also used to extrapolate $P(\etac|\zc,\zp)$ to
different polymer fugacities $\zp'$ to obtain estimates of
$P(\etac|\zc,\zp')$. To this end, also the distribution of the number of
polymers must be recorded for each window. Since our implementation of
successive umbrella sampling only puts a bias on the number of colloids,
and not on the polymers, this extrapolation will introduce an error. The
accuracy of the estimated distribution deteriorates when the range $\zp' -
\zp$ over which one extrapolates becomes larger. Fortunately, since we are
primarily interested in the behavior close to the critical point, the
range need not be large and the error introduced by histogram reweighting
is small~\cite{deutsch1992a}. More importantly, the error can easily be
checked for as will be shown later.

Note that in terms of histogram reweighting, the AO model is extremely
convenient. Since the histograms that need to be maintained involve integer data
only (namely numbers of particles) the problem of choosing a bin size for
example does not occur.

\section{Extracting observables}
\label{sec:obs}

The coexistence distribution $\Pc$ is a powerful quantity because a number of
important physical observables can be extracted from it. For instance, the
average packing fraction of the colloidal vapor can be written as:
\begin{equation}
  \etacv = 2 \int_0^\avg{\etac} \etac \Pc {\rm d} \etac,
\end{equation}
and a similar expression holds for the average packing fraction of the colloidal
liquid:
\begin{equation}
 \etacl = 2 \int_\avg{\etac}^\infty \etac \Pc {\rm d} \etac,
\end{equation}
with $\avg{\etac}$ given by \eq{eq:avg}. In these and following equations, the
normalization condition of \eq{eq:norm} is assumed (this also explains the
origin of the factors of two in the above two equations). 

For the AO model, we define in the two--phase region half the difference
in colloid packing fraction between the vapor and liquid phase as order
parameter. The order parameter is denoted $M_{\rm c}$:
\begin{equation}
\label{eq:order}
 M_{\rm c} \equiv \frac{\etacl-\etacv}{2} 
 = \int_0^\infty | \etac - \avg{\etac} | \Pc {\rm d} \etac,
\end{equation}
and the coexistence diameter (or rectilinear diameter) is given by:
\begin{equation}
 D_{\rm c} \equiv \frac{\etacl + \etacv}{2},
\end{equation}
where the subscripts ``c'' emphasize that the definitions are expressed in
terms of the colloid packing fraction, and not the polymer packing
fraction.

In a similar way, we identify the ``concentration susceptibility'' or
compressibility as the variance of the peaks in $\Pc$~\cite{das2003a}. In
the two--phase region ($\etapr > \etaprcr$), we define the susceptibility
of the colloidal vapor as:
\begin{equation}
\label{eq:susv}  
\chi_{\rm c}^{\rm v} = V \left[ \left( 
  2 \int_0^\avg{\etac} \etac^2 P(\etac) {\rm d} \etac \right) -
  (\etacv)^2 \right],
\end{equation}
and the susceptibility of the colloidal liquid as:
\begin{equation}
\label{eq:susl}  
  \chi_{\rm c}^{\rm l} = V \left[ \left( 
  2 \int_\avg{\etac}^\infty \etac^2 P(\etac) {\rm d} \etac \right) -
  (\etacl)^2 \right],
\end{equation}
where the factors of two are again consequence of the normalization
condition of \eq{eq:norm}. Note also the presence of the volume $V$ in the
above definitions. In the one--phase region ($\etapr < \etaprcr$), the
distribution $\Pc$ will loose its bimodal structure and become single
peaked. In this regime the correct definition for the susceptibility
reads~\cite{deutsch1992a, binder2001a}:
\begin{equation}
\label{eq:sus1}
  \chi_{\rm c} = V \left[ \left(
  \int_0^\infty \etac^2 P(\etac) {\rm d} \etac \right) -
  \avg{\etac}^2 \right],
\end{equation}
with $\avg{\etac}$ given by \eq{eq:avg}. 

The above definitions are easily modified to define the corresponding
observables for the polymer phases (simply replace the subscript ``c'' by
``p''). Since our simulations also store the polymer histograms, coexisting
packing fractions and susceptibilities can be calculated for these phases as
well. Note that the above definitions are additionally attractive from a
numerical point of view because the integrations over $\etac$ tend to average
out the statistical fluctuations that may be present in $\Pc$. 

\begin{figure}
\begin{center}
\includegraphics[clip=,width=\figwidth]{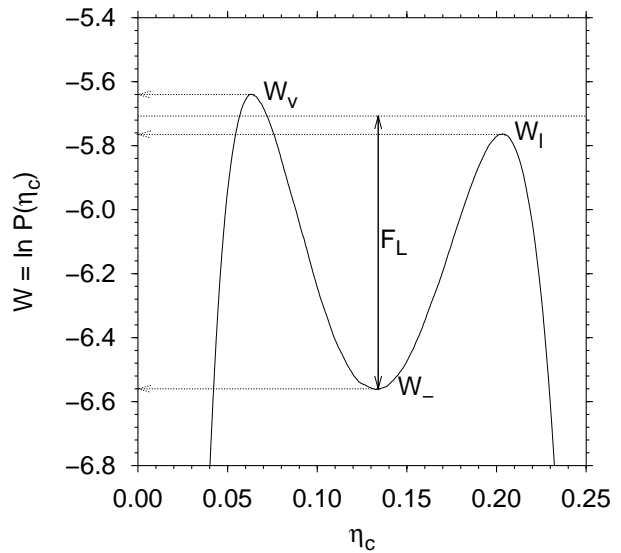}
\caption{~\label{fig:bimodal}Logarithm of the probability distribution $W \equiv
\ln P(\etac)$ for an AO model with $q=0.8$ and $\etapr=0.765$ at coexistence.
The data were obtained in a cubic simulation cell with edge $L=21.0$ and using
periodic boundary conditions. The peak at low $\etac$ corresponds to the
colloidal vapor, the peak at high $\etac$ to the colloidal liquid. $\wv$ ($\wl$)
denotes the maximum value of the colloidal vapor (liquid) peak. $W_-$ is the
minimum value of $W$ between the two peaks. $F_L$ corresponds to the free energy
barrier separating the vapor phase from the liquid phase.}
\end{center}
\end{figure}

The interfacial tension is extracted from the logarithm of the probability
distribution: $W \equiv \ln P(\etac)$. Since $W$ corresponds to the free
energy of the system, the height of the peaks in $W$ may be identified as
the free energy barrier separating the colloidal vapor from the colloidal
liquid~\cite{binder1982a}. In \fig{fig:bimodal} the barrier is marked
$F_L$, where the subscript $L$ emphasizes that the data stem from a finite
simulation box of size $L$. In practice, $F_L$ is extracted from $W$ via:
\begin{equation}
\label{eq:fl}
  F_L = W_+ - W_-, 
\end{equation}
where $W_+$ is the average of $W$ in the peaks: $W_+ = (\wv + \wl)/2$, and
$W_-$ the value of $W$ at the minimum between the peaks (the symbols $\wv$ and
$\wl$ are defined in \fig{fig:bimodal}). The corresponding interfacial 
tension for the finite system reads~\cite{binder1982a}:
\begin{equation}
\label{eq:st}
  \sigma_L = F_L/(2L^2),
\end{equation}
where the factor of two stems from the use of periodic boundary conditions which
yield the formation of two interfaces in the system.

\section{Critical behavior}

Essential in the study of critical phenomena is some measure of distance from
the critical point. As measure of distance we use in this work the parameter:
\begin{equation}
  t = (\etapr / \etaprcr - 1),
\end{equation}
which is positive in the two--phase region ($\etapr > \etaprcr$) and
negative in the one--phase region ($\etapr < \etaprcr$). This implies that 
in the one--phase region we must use $-t$.

When the critical point is approached from the two--phase region, $M$ and
$\sigma$ vanish precisely, while $\chi$ diverges. Common symbols have been
established to denote the critical exponents and critical amplitudes of the
associated power laws:
\begin{equation}
  M = B t^\beta, \hspace{3mm}
  \sigma = \sigma_0 t^{2\nu}, \hspace{3mm}
  \chi = (\Gamma^-) t^{-\gamma},
\end{equation}
where hyperscaling has been assumed (valid for systems that belong to the 
3D Ising universality class).

When the critical point is approached from the other side, namely the one--phase
region, $M$ and $\sigma$ remain zero, while $\chi$ diverges:
\begin{equation}
  \chi = \Gamma^+ (-t)^{-\gamma},
\end{equation}
but with a different critical amplitude $\Gamma^+$.

The critical behavior of the coexistence diameter $D$ in the two--phase 
region is given by~\cite{rehr1973a, nakata1978a}:
\begin{equation}
\label{eq:mstar}
 D - X_{\rm cr} = A t^{1-\alpha},
\end{equation}
with $X_{\rm cr}$ the packing fraction at criticality, given by $\etaccr$
($\etapcr$) in case of the colloid (polymer) coexistence diameter. In the
one--phase region $D$ is not well defined because the distinction between
a colloidal vapor and liquid is then no longer possible.

For the 3D Ising universality class, the critical exponents are given by:
\begin{eqnarray}
\label{eq:3di}
  \beta=0.324, \hspace{5mm}
  \gamma=1.239, \\
  \nu=0.629,  \hspace{5mm}
  \alpha=0.113 \nonumber.
\end{eqnarray}
Also certain combinations of the critical amplitudes are
universal~\cite{stauffer1972, privman1991a, pelissetto2002}. Of importance
in this work are the ratios:
\begin{eqnarray}
  \label{eq:amp1} 
  U_2 &=& \Gamma^+ / \Gamma^- 
    \approx 4.76 \pm 0.24, \\
  \label{eq:amp2}
  w^2 R_\sigma^{3/2} &=& \frac{\sigma_0^{3/2} \Gamma^-}{B^2} 
    \approx 0.13 \pm 0.04, \\
  \label{eq:amp3}
  \frac{(R_\sigma^+)^{3/2}}{Q_c} &=& \frac{\sigma_0^{3/2} \Gamma^+}{B^2} 
    \approx 0.71 \pm 0.13, 
\end{eqnarray}
with values taken from \olcite{pelissetto2002}, where $k_{\rm B} T$ is
used as the unit of energy. The ranges represent the spread in different
estimates obtained from experiment, simulation, and theory. Note that the
ranges are quite substantial, indicating that the determination of
critical amplitudes is still a challenge.

\section{Finite Size Scaling}
\label{sec:fss}

In a computer simulation, the critical power laws given in the previous
section cannot be observed directly. As explained before, this is related
to the correlation length which diverges at the critical point and hence
cannot be captured in a finite system of size $L$. However, it is possible
to perform several simulations at different system sizes and use the
predictions of finite size scaling theory to extrapolate to the
thermodynamic limit. Reviews of the subject are abundant in the
literature~\cite{deutsch1992a, binder2001a, lb_book}. In this section, we
will therefore be brief and only reproduce the equations required for our
analysis.

\subsection{Finite size scaling of $M$, $\chi$ and $D$}

According to finite size scaling theory, the order parameter $M_L$
obtained in a finite system of linear dimension $L$ close to the critical
point shows a systematic $L$ dependence that can be written 
as~\cite{lb_book}:
\begin{equation}
  M_L = L^{-\beta/\nu} {\cal M}^0 (t L^{1/\nu}),
\end{equation}  
with $t$ the distance from the critical point, critical exponents
$\{\beta,\nu\}$ and ${\cal M}^0$ some function independent of system size
(${\cal M}^0$ is called a scaling function). The critical exponents
appropriate for the AO model are the 3D Ising exponents listed in
\eq{eq:3di}. Since ${\cal M}^0$ is system size-independent, this implies
that plots of $L^{\beta/\nu} M_L$ versus $t L^{1/\nu}$ should all collapse
onto one master curve, provided the correct values of the critical polymer
fugacity and the critical exponents are used. Such scaling
plots~\cite{deutsch1992a} thus give an indication of whether the assumed
universality class is indeed correct. Moreover, for large $t L^{1/\nu}$
(but still within the critical region of course) such plots should
approach the power law behavior of the thermodynamic limit. This is best
visualized if a double logarithmic scale in the scaling plot is used. The
data should then approach a straight line, with slope~$\beta$ and
intercept equal to the critical amplitude $B$.

The scaling behavior of the susceptibility is analogously given by:
\begin{equation}
  \chi_L = L^{\gamma/\nu} \chi^0 (t L^{1/\nu}). 
\end{equation}
In this case, the appropriate quantities for the scaling plot are
$L^{-\gamma/\nu} \chi_L$ and $t L^{1/\nu}$. On a double logarithmic scale,
the data should approach a straight line, with slope~$-\gamma$ and
intercept equal to the critical amplitude $\Gamma^-$ or $\Gamma^+$
(depending on whether the critical point is approached from the one--phase
or the two--phase region).

By presenting the simulation data in the form of scaling plots,
measurements of the critical amplitudes become possible. The accuracy of
the method increases when larger system sizes are used. In order for
finite size scaling theory to apply, it is important that the simulations
are performed in the so--called scaling regime. At which system size $L$
the scaling regime begins, varies from system to system and is {\it a
priori} not known. If, however, the considered system sizes are too small,
it will show in the corresponding scaling plots as systematic deviations
from any data collapse onto a master curve. Alternatively, one could
consider corrections to finite size scaling theory for these smaller
systems~\cite{luijten2002a}.

In principle, a finite size scaling analysis of the coexistence diameter
$D$ is also possible. In this case, the appropriate scaling plot is $(D -
X_{\rm cr}) L^{(1-\alpha)/\nu}$ expressed as a function of $t L^{1/\nu}$.
In practice, however, it is difficult to distinguish in simulation data
the term $t^{1-\alpha}$ in \eq{eq:mstar} from the next order term in the
series expansion (which would be linear in $t$) because the critical
exponent $\alpha$ is rather small. The accuracy of such scaling plots is
therefore usually rather poor.

\subsection{Finite size scaling of $\sigma$}

The interfacial tension in the thermodynamic limit $\sigma_\infty$ (in $d$
dimensions) is related to the free energy barrier of the finite system
$F_L$ via~\cite{binder1982a}:
\begin{equation}
 \exp(F_L) = A L^x \exp(2 L^{d-1} \sigma_\infty),
\end{equation}
where trivial factors of $k_B T$ have been dropped. Taking the logarithm 
on both sides, the above equation can be written as:
\begin{equation}
\label{eq:fss}
  \sigma_L = \sigma_\infty + \frac{ x \ln L }{ 2L^{d-1} }
  + \frac{ \ln A }{ 2L^{d-1} },
\end{equation}
with $\sigma_L$ the interfacial tension of the finite system given by
\eq{eq:st} and constants $\{x,A\}$ that are generally not known. While it
is not possible to measure $\sigma_\infty$ directly, it is possible to
measure the interfacial tension of the finite system $\sigma_L$ for
several system sizes $L$, and then use the above equation to extrapolate
to the thermodynamic limit.

One attractive feature of \eq{eq:fss} is that it does not depend on the
critical exponent $\nu$ and can therefore be applied without prior
knowledge of the universality class of the system. In other words, it can
be used to measure $\nu$ as is demonstrated in \olcite{potoff2000a} for
the Lennard-Jones fluid. For the AO model, the universality class is
already known~\cite{vink2004a, vink2004b}. In this case, \eq{eq:fss} still
provides a powerful consistency check: if the universality class of the AO
model is indeed 3D Ising, it should be possible to extract the exponent
$\nu$ from the simulation data.

The issue here is how to perform the extrapolations in
practice~\cite{mon1988a, berg1993a, hunter1995a}. Ideally, the simulation
data for the different system sizes should be extrapolated to the
thermodynamic limit using two--parameter fits in the variables
$\ln(L)/L^{d-1}$ and $1/L^{d-1}$. This approach, however, may have
numerical problems associated with it. Typically, only data over a
relatively small range of different system sizes is available (this is
certainly the case for a non--trivial mixture like the AO model). It will
be difficult to separate the $\ln(L)/L^{d-1}$ term accurately from the
$1/L^{d-1}$ term over such a small range. Therefore, this extrapolation
scheme will likely lead to poor precision.

Alternatively, one can argue that for small $L$ it may occur that $|x \ln
L| < |\ln A|$, in which case an extrapolation in the single variable
$1/L^{d-1}$ is most appropriate, whereas for large $L$ the single variable
$\ln(L)/L^{d-1}$ is the better choice~\cite{binder1982a}. In
\olcite{potoff2000a}, for example, the extrapolations are performed in the
single variable $\ln(L)/L^{d-1}$. Since it is {\it a priori} not clear
which of the above extrapolation methods is the most accurate, all are
investigated in this work.

\section{Results}

The following results stem from simulations of the AO model with $q=0.8$
performed in cubic boxes with edge length $L$ and using periodic boundary
conditions. The dimensionality of the simulations is $d=3$. In order to
apply finite size scaling, the following system sizes are considered:
$L_1=15.5$, $L_2=16.7$, $L_3=17.7$ and $L_4=21.0$. For each system size,
the coexistence probability $\Pc$ is measured accurately at one value of
$\etapr$ chosen in the vicinity of the critical point. Histogram
reweighting is used to extrapolate $\Pc$ to other values of $\etapr$. We
also performed a number of shorter simulations to explicitly measure $\Pc$
at different values of $\etapr$. The results of these simulations were
used to check the consistency and accuracy of the extrapolated
distributions. The quality of our data is such that extrapolations over
the range $\etapr \approx 0.72$ to $\etapr \approx 0.83$ can be carried
out reliably.

\subsection{Order parameter and binodal}

\begin{figure}
\begin{center}
\includegraphics[clip=,width=\figwidth]{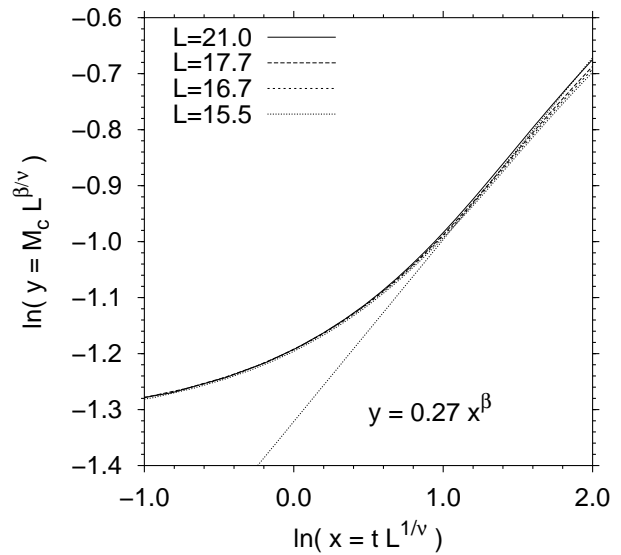}    

\caption{~\label{fig2}3D Ising scaling plot for the order parameter of the
colloids $M_{\rm c}$ given by \eq{eq:order} for the AO model with $q=0.8$
in the two--phase region. The choice of axis is explained in the text. The
critical amplitude is extracted by means of a fit to the tail of the data,
see also \tab{tab:crit}.}

\end{center}
\end{figure}

\begin{figure}
\begin{center}
\includegraphics[clip=,width=\figwidth]{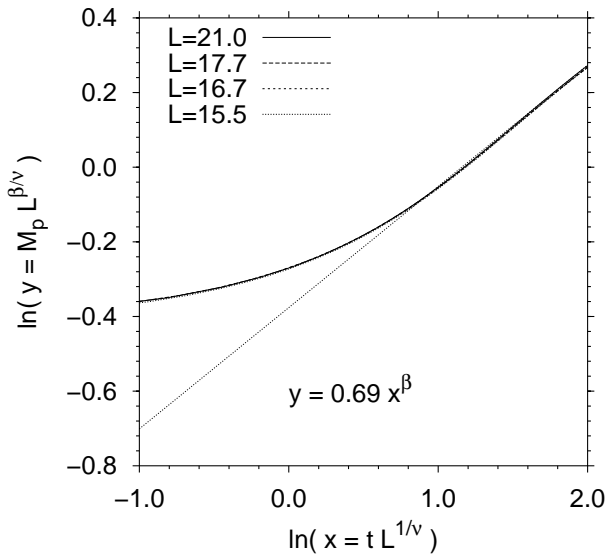}    
\caption{~\label{fig3} The analogue of \fig{fig2} for the order 
parameter of the polymers $M_{\rm p}$.}
\end{center}
\end{figure}

The critical behavior of the order parameter of the colloids $M_{\rm c}$
is analyzed in the scaling plot of \fig{fig2}. The collapse of the
simulation data from the different system sizes onto one master curve is
clearly visible. The scaling plot for the order parameter of the polymers
$M_{\rm p}$ is shown in \fig{fig3}, which again demonstrates the collapse
onto one master curve. In these figures, the critical polymer fugacity
$\etaprcr$ was used as a free parameter and tuned until the best collapse
occurred. By performing a linear least squares fit to the tails of the
master curves, the critical amplitudes can be obtained. The corresponding
critical power laws are given in \tab{tab:crit}, which also lists the
value of $\etaprcr$ that was used in the scaling plots. Naturally, this
value of $\etaprcr$ should agree with the previous estimate listed in
\eq{eq:old}. The critical amplitudes obtained from the fits are sensitive
to the range over which the fit is performed: the variation is used as a
measure for the error in \tab{tab:crit}.

\begin{figure}
\begin{center}
\includegraphics[clip=,width=\figwidth]{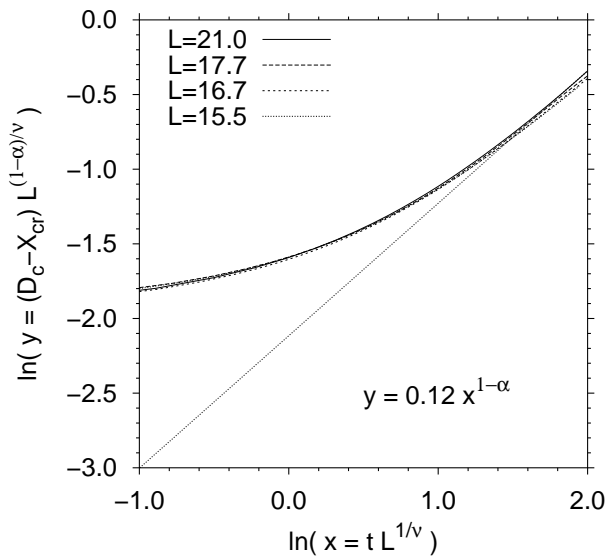}
\caption{~\label{fig4}Scaling plot of the coexistence diameter of the
colloids $D_{\rm c}$. The value of $X_{\rm cr} = \etaccr$ used in this
plot was taken from \eq{eq:old}. Allowing variations in $X_{\rm cr}$ did
not improve the collapse of the data.}
\end{center}
\end{figure}

\begin{figure}
\begin{center}
\includegraphics[clip=,width=\figwidth]{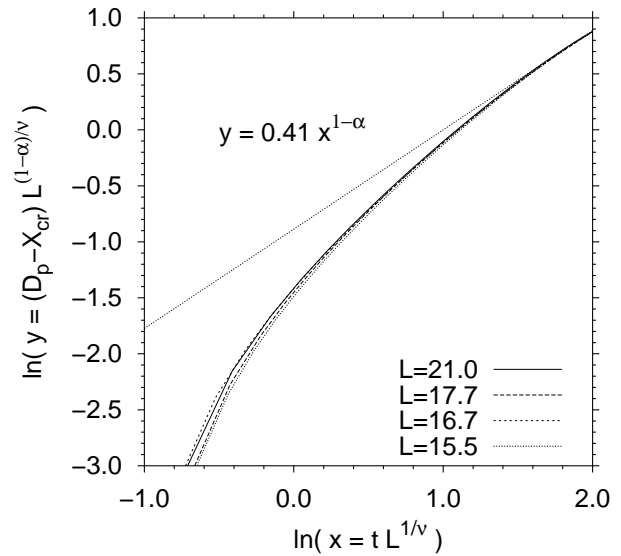}
\caption{~\label{fig5}The analogue of \fig{fig4} for the coexistence
diameter of the polymers $D_{\rm p}$. $X_{\rm cr} = \etapcr$ was taken
from \eq{eq:old}.}
\end{center}
\end{figure}

The critical behavior of the coexistence diameter of the colloids $D_{\rm
c}$ is presented in \fig{fig4}. In this case, the best collapse of the
data occurs at $\etaprcr=0.771$ which is not in agreement with
\eq{eq:old}. Some discrepancy is to be expected though because the
singularity in the coexistence diameter is very weak, which makes it hard
to discern it from simulation data. The behavior of the data in the tails,
however, seems rather well described by the exponent $1-\alpha$. The
corresponding scaling plot for the coexistence diameter of the polymers
$D_{\rm p}$ is shown in \fig{fig5}. In this case, the curvature of the
data for small $t$ seems in error. Therefore, we conclude that our
simulation data is not accurate enough to reliably extract the critical
behavior of the coexistence diameter. The power laws for $D_{\rm c}$ and
$D_{\rm p}$, given for completeness in \tab{tab:crit}, must therefore be
treated with some care.

\begin{figure}
\begin{center}
\includegraphics[clip=,width=\figwidth]{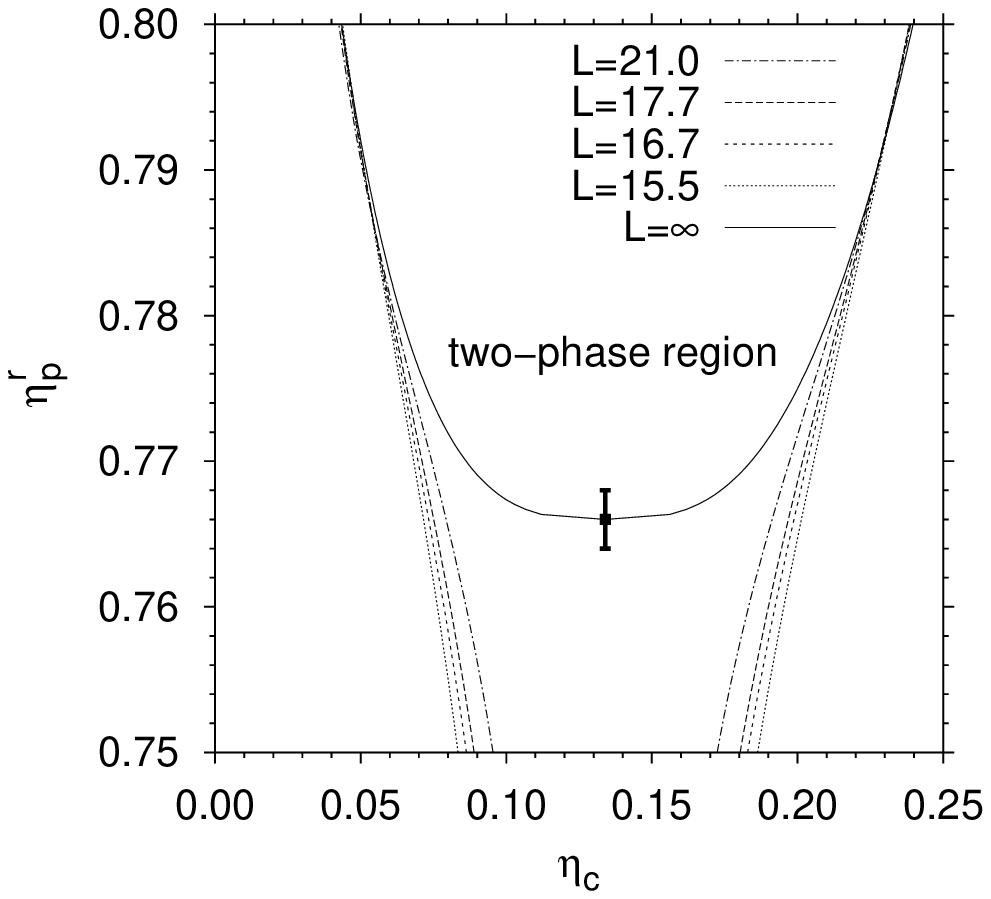}
\caption{~\label{fig6} Phase diagram of the AO model in reservoir
representation. The solid curve ($L=\infty$) shows the binodal in the
thermodynamic limit obtained using the power laws for $M_{\rm c}$ and
$D_{\rm c}$ listed in \tab{tab:crit} (here $\etaprcr=0.766$ was used).  
The dashed/open curves are raw simulation data for various finite system
sizes $L$ as indicated. The bar marks the location of the critical point
given by \eq{eq:old}.}
\end{center}
\end{figure}

\begin{figure}
\begin{center}
\includegraphics[clip=,width=\figwidth]{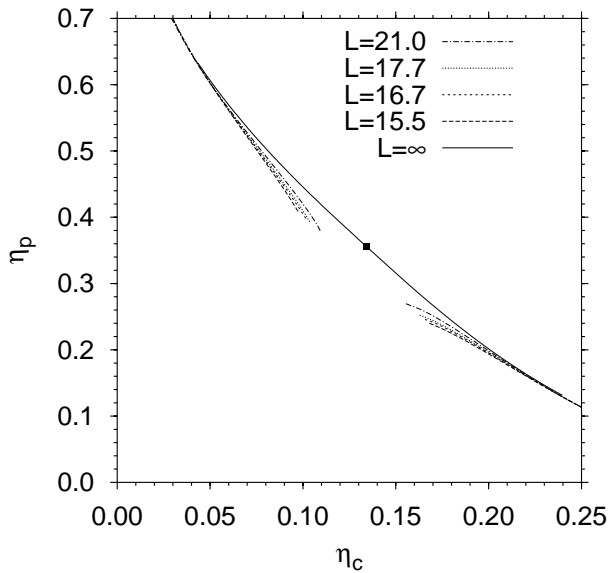}
\caption{~\label{fig7}Phase diagram of the AO model in system
representation. The solid curve ($L=\infty$) shows the binodal in the
thermodynamic limit obtained using finite size scaling (here
$\etaprcr=0.766$ was used). The dashed/open curves are raw simulation data
for various finite system sizes $L$ as indicated. The square marks the
location of the critical point given by \eq{eq:old}.}
\end{center}
\end{figure}

By combining the expressions for $M_{\rm c}$ and $D_{\rm c}$ in
\tab{tab:crit}, the colloid packing fractions of the vapor and liquid
phase ($\etacv$ and $\etacl$) can be written as functions of $\etapr$.
These expressions, which are valid close to the critical point in the
thermodynamic limit, describe the binodal of the AO model in reservoir
representation. The resulting binodal is shown in \fig{fig6}, together
with the raw simulation data from the various system sizes. The figure
clearly shows the familiar finite--size deviation of the simulation data
close to the critical point~\cite{deutsch1992a}. If also the critical
expressions for $M_{\rm p}$ and $D_{\rm p}$ are used, $\etapr$ can be
eliminated altogether to yield the binodal in the experimentally more
relevant $(\etac,\etap)$ or system representation. The resulting plot is
shown in \fig{fig7}, together with the raw simulation data.

\subsection{Susceptibility}

\begin{figure}
\begin{center}
\includegraphics[clip=,width=\figwidth]{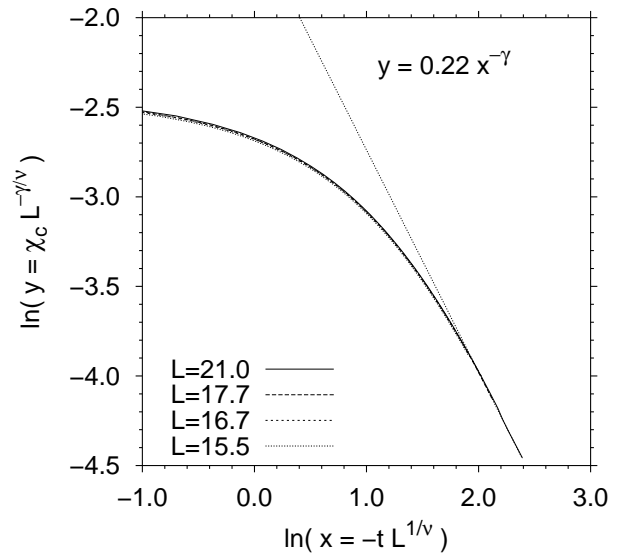}
\caption{~\label{fig8}Scaling plot of the susceptibility of the colloidal
phase $\chi_{\rm c}$ in the one--phase region. In this case \eq{eq:sus1}
is used to measure $\chi_{\rm c}$.}
\end{center}
\end{figure}

Next, we consider the critical behavior of the susceptibility. \fig{fig8} shows
the scaling plot of the susceptibility of the colloidal phase $\chi_{\rm c}$ in
the one--phase region. The critical power law extracted from this plot is listed
in \tab{tab:crit}. The scaling plot for the susceptibility of the polymer phase
$\chi_{\rm p}$ in the one--phase region is qualitatively similar and not shown.
Instead, only the critical power law is given in \tab{tab:crit}.

Measurements of the susceptibility in the two--phase region are prone to a
number of potential numerical pitfalls. Since the susceptibility is a second
order moment of the distribution $\Pc$, it is generally more sensitive to
statistical errors than first order moments like the order parameter. For
asymmetric systems like the AO model, the additional problem arises that the
statistics in the liquid peak of $\Pc$ are systematically better than in the
vapor peak: simply because the liquid peak contains more colloids. These
problems will of course vanish with increasing system size. However, for a
non-trivial system like the AO model, the system sizes that can be handled today
are unfortunately still in the regime where these subtleties come into play.

\begin{figure}
\begin{center}
\includegraphics[clip=,width=\figwidth]{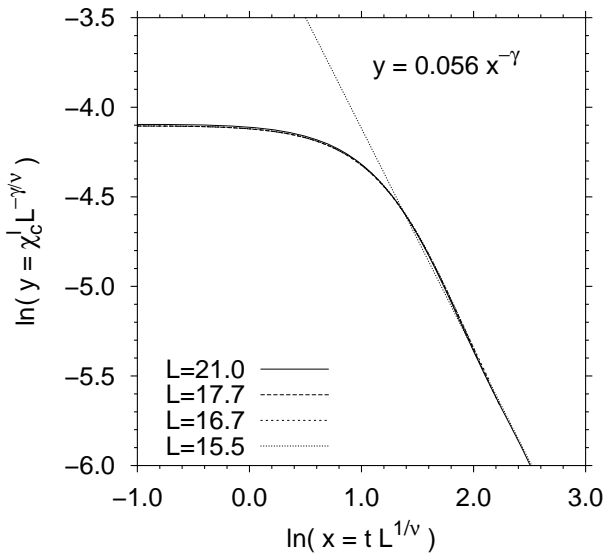}
\caption{~\label{fig9}Scaling plot of the susceptibility of the colloidal
liquid $\chi_{\rm c}^{\rm l}$ given by \eq{eq:susl} in the two--phase
region.}
\end{center}
\end{figure}

\begin{figure}
\begin{center}
\includegraphics[clip=,width=\figwidth]{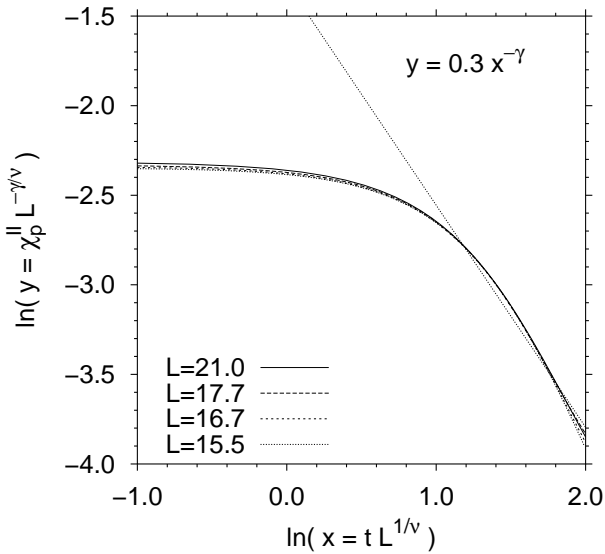}
\caption{~\label{fig10}Scaling plot of the average susceptibility of the
two polymer phases $\chi_{\rm p}^{\rm II}$ in the two phase region.}
\end{center}
\end{figure}

In case of the colloidal vapor and liquid in the two--phase region, the
relevant susceptibilities are $\chi_{\rm c}^{\rm v}$ given by \eq{eq:susv}
and $\chi_{\rm c}^{\rm l}$ given by \eq{eq:susl}. In principle, close to
the critical point, the susceptibility should be the same in both phases.
In practice, there will be differences due to the numerical difficulties
outlined above. Two procedures are now conceivable: (1) use the average of
$\chi_{\rm c}^{\rm v}$ and $\chi_{\rm c}^{\rm l}$ as best estimate of the
susceptibility, or (2) use only $\chi_{\rm c}^{\rm l}$. Which of these
procedures is the best needs to be checked with the data available. In our
case, the susceptibility of the colloidal phase in the two--phase region
was best described using only $\chi_{\rm c}^{\rm l}$. The resulting
scaling plot is shown in \fig{fig9} and the corresponding power law in
\tab{tab:crit}. The collapse of the data is clearly visible, but unlike
\fig{fig8}, the slope of the data for large $t$ is not quite~$-\gamma$. A
less satisfactory fit is expected though, because the statistics in
\fig{fig9} are significantly worse compared to \fig{fig8} since only half
the data is used.

The susceptibility measurements of the polymer phases in the two--phase
region were most accurate if the average $\chi_{\rm p}^{\rm II} \equiv
(\chi_{\rm p}^{\rm l} + \chi_{\rm p}^{\rm v})/2$ was used. The scaling
plot of $\chi_{\rm p}^{\rm II}$ is shown in \fig{fig10} and the
corresponding power law is listed in \tab{tab:crit}. As in \fig{fig9}, the
data collapse is clearly demonstrated, but the slope~$-\gamma$ is not
quite reproduced.

\begin{figure}
\begin{center}
\includegraphics[clip=,width=\figwidth]{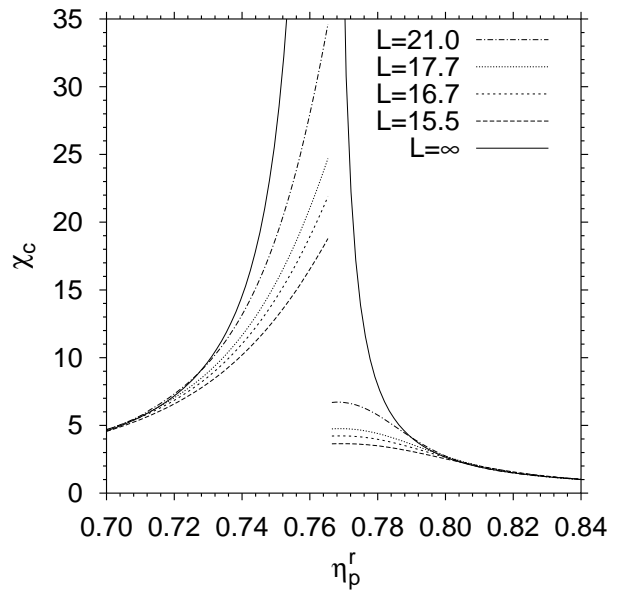}
\caption{~\label{fig11}Susceptibility of the colloids $\chi_{\rm c}$
across the phase transition. The solid curve shows the susceptibility in
the thermodynamic limit given by the critical power laws of \tab{tab:crit}
with $\etaprcr=0.766$. The remaining curves show the raw simulation data
of the various system sizes $L$.}
\end{center}  
\end{figure}

In \fig{fig11} we plot the susceptibility of the colloids as function of
$\etapr$ in the vicinity of the critical point. Shown are the raw
simulation data as well as the critical power laws of \tab{tab:crit}.
Clearly demonstrated is the familiar finite--size rounding of the raw
simulation data in the vicinity of the singularity~\cite{deutsch1992a}.

\subsection{Interfacial tension}

\begin{figure}
\begin{center}
\includegraphics[clip=,width=\figwidth]{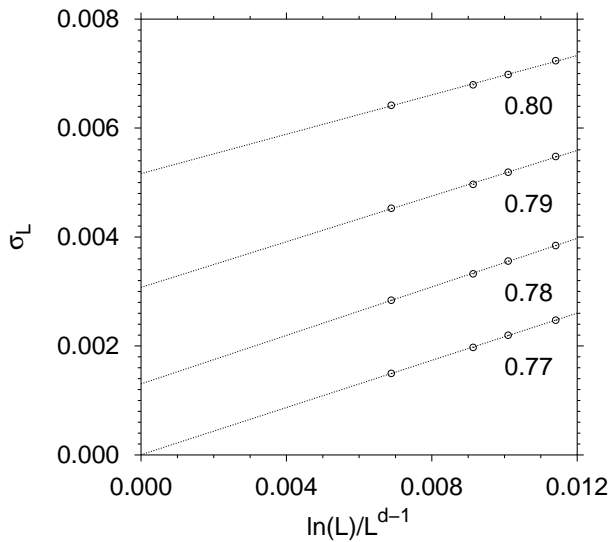}
\caption{~\label{fig12}Finite size extrapolation of the interfacial
tension in $\ln(L)/L^{d-1}$. This extrapolation scheme corresponds to the
assumption $\ln A=0$ in \eq{eq:fss}. Shown in the above plot is the
interfacial tension of the finite system $\sigma_L$ as function of
$\ln(L)/L^{d-1}$ for various values of $\etapr$ as indicated. The straight
lines are linear least squares fits to the data. The intercept of these
lines with the ordinate yields an estimate for the interfacial tension
$\sigma_\infty$ in the thermodynamic limit.}
\end{center}
\end{figure}

\begin{figure}
\begin{center}
\includegraphics[clip=,width=\figwidth]{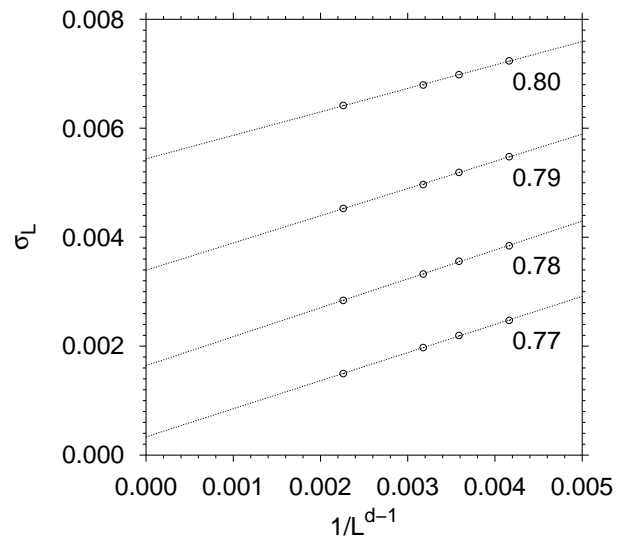}
\caption{~\label{fig13}The analogue of \fig{fig12} but using $1/L^{d-1}$
as scaling variable instead. This extrapolation scheme corresponds to the
assumption $x=0$ in \eq{eq:fss}.}
\end{center}
\end{figure}

As mentioned before, the finite size extrapolation of the interfacial
tension is less straightforward and various methods can be used. Since the
number of different system sizes considered by us is rather small,
multi-parameter fits such as the ones investigated in \olcite{berg1993a}
are out of the question. Instead, we investigate single parameter fits
only, which in this case essentially means choosing $\ln(L)/L^{d-1}$ or
$1/L^{d-1}$ as scaling variable. The results of both extrapolation
procedures for a number of different $\etapr$ are summarized in
\fig{fig12} and \fig{fig13}. The extrapolations produce meaningful
estimates up to $\etapr \approx 0.80$. For higher values of $\etapr$, one
leaves the critical regime and \eq{eq:fss} breaks down. In this regime,
the interfacial tension gradually becomes system size independent.

\begin{figure}
\begin{center}
\includegraphics[clip=,width=\figwidth]{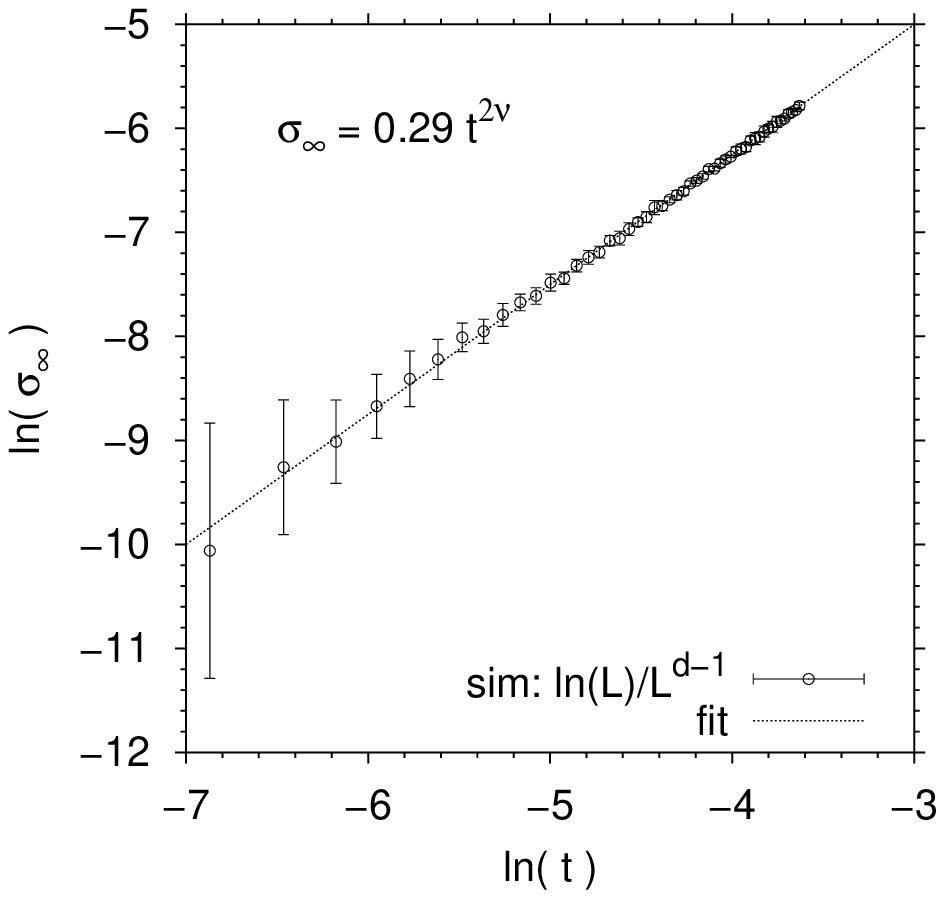}
\caption{~\label{fig14}Interfacial tension in the thermodynamic limit
$\sigma_\infty$ as function of $t$ on a double logarithmic scale, where
$\ln(L)/L^{d-1}$ was used as scaling variable. The best collapse onto a straight
line is observed at $\etaprcr=0.7696$, which is not in agreement with
\eq{eq:old}. The slope of the line yields the critical exponent of the
interfacial tension, for which we find $2\nu=1.25 \pm 0.01$.}
\end{center}
\end{figure}

\begin{figure}
\begin{center}
\includegraphics[clip=,width=\figwidth]{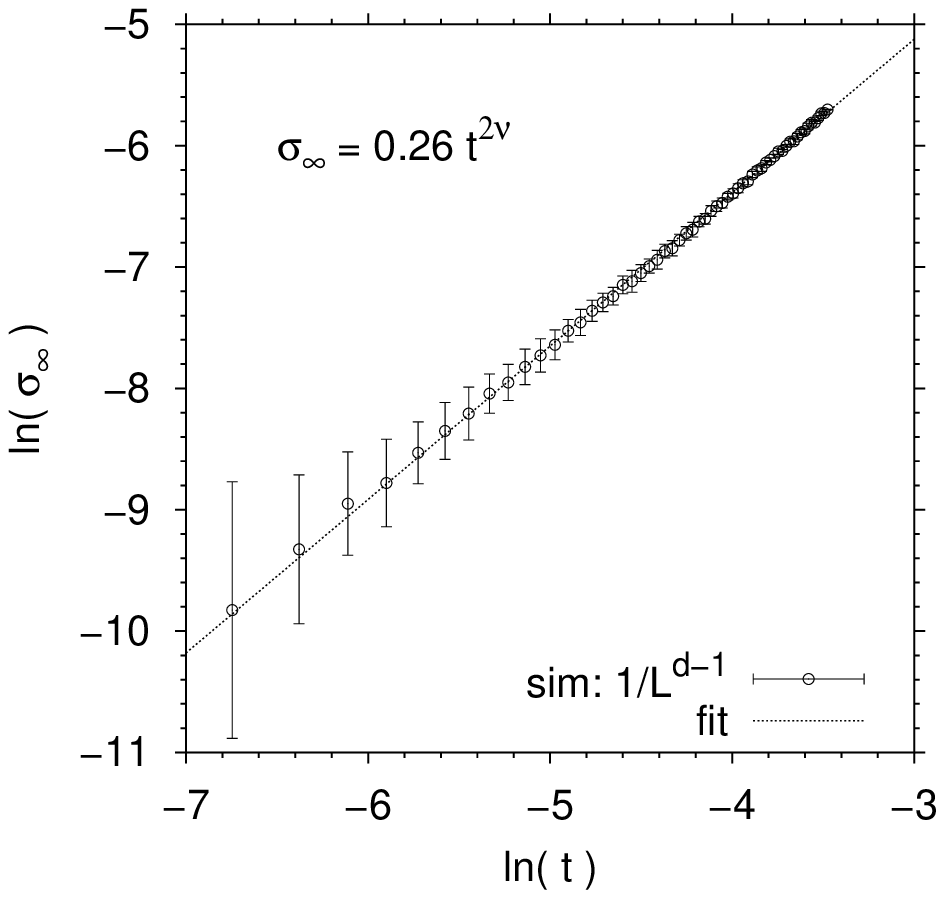}
\caption{~\label{fig15}The analogue of \fig{fig14} in which $1/L^{d-1}$ is used
as scaling variable instead. The best collapse occurs \mbox{at
$\etaprcr=0.7661$}, which is in excellent agreement with \eq{eq:old}. For 
the critical exponent we obtain $2\nu=1.26 \pm 0.01$ and for the critical 
amplitude $\sigma_0=0.26 \pm 0.02$.}
\end{center}
\end{figure}

The fits in \fig{fig12} and \fig{fig13} seem equally accurate, and based
on these figures alone we cannot reject one extrapolation method in favor
of the other. The issue is resolved when the interfacial tension
$\sigma_\infty$ in the thermodynamic limit itself is considered. In this
case, the critical power law $\sigma_\infty= \sigma_0 t^{2\nu}$ is valid
which implies that plots of $\sigma_\infty$ versus $t$ on double
logarithmic scales should collapse onto straight lines, provided the
correct value of $\etaprcr$ in $t$ is used. Note that $\etaprcr$ is the
only free parameter: the critical exponent $\nu$ follows automatically
from the slope of the line. The resulting plots are shown in \fig{fig14}
and \fig{fig15}, in which $\ln(L)/L^{d-1}$ and $1/L^{d-1}$ were used as
scaling variable, respectively, and $\etaprcr$ in each plot was tuned
until the best collapse occurred. In these figures, measurements of the 
interfacial tension up to $\etapr=0.79$ were used.

\begin{table}[b]
\caption{\label{tab:crit}Summary of the critical behavior of the AO model
with $q=0.8$. Shown are the critical power laws in the one--phase and
two--phase regions of various physical quantities obtained in scaling
plots. Listed in the last column is the value of $\etaprcr$ at which the
best collapse in the corresponding scaling plot was observed (provided
here to give an indication of the consistency of our results). The symbols
$M$, $D$, $\chi$ and $\sigma$ are defined in section~\ref{sec:obs}, while
the critical exponents are listed in \eq{eq:3di}.}
\vspace{2mm}
\begin{ruledtabular}
\begin{tabular}{c|c|c|c} 
  & one--phase region  & two--phase  region & $\etaprcr$ \\ \hline
$M_{\rm c}$ & -- & $(0.27 \pm 0.02) t^\beta$ & 0.765 \\
$M_{\rm p}$ & -- & $(0.69 \pm 0.01) t^\beta$ & 0.765 \\ \hline

$D_{\rm c}-\etaccr$ & -- & $(0.12 \pm 0.01) t^{1-\alpha}$ & 0.771 \\
$D_{\rm p}-\etapcr$ & -- & $(0.41 \pm 0.01) t^{1-\alpha}$ & 0.766 \\ \hline

$\chi_{\rm c}$ & $(0.22 \pm 0.03) (-t)^{-\gamma}$ & & 0.766 \\
               & & $(0.056 \pm 0.005) t^{-\gamma}$ & 0.766 \\ \hline

$\chi_{\rm p}$ & $(1.24 \pm 0.08) (-t)^{-\gamma}$ & & 0.765  \\
               & & $(0.3 \pm 0.1) t^{-\gamma}$  & 0.768  \\ \hline

$\sigma$ & -- & $(0.26 \pm 0.02) t^{2\nu}$ & 0.766 \\
\end{tabular}
\end{ruledtabular}
\end{table}

One important observation is that both data sets in \fig{fig14} and
\fig{fig15} accurately reproduce the expected slope $2\nu \approx 1.26$
corresponding to 3D Ising behavior. However, if $\ln(L)/L^{d-1}$ is used
as scaling variable, the best collapse is obtained at $\etaprcr=0.7696$,
which is not in agreement with the previous estimate of \eq{eq:old}. On
the other hand, if $1/L^{d-1}$ is used, the best collapse is observed at
$\etaprcr=0.7661$, which is in excellent agreement with \eq{eq:old}.
Therefore, we conclude that $1/L^{d-1}$ is the appropriate scaling
variable for our problem. The corresponding power law obtained from
\fig{fig15} is listed in \tab{tab:crit}. The effect of using the incorrect
scaling variable seems to be that the critical point is not correctly
estimated, while the critical exponent is less affected. It thus seems
wise practice to always compare the critical point obtained from finite
size extrapolations of the interfacial tension to some other independent
estimate (this estimate could for instance be obtained using the cumulant
intersection method). 

It is worth noting that Berg {\it et al.}~in \olcite{berg1993a} also
conclude that the extrapolation of the interfacial tension is most
consistent if $x=0$ is assumed in \eq{eq:fss}. In retrospect, it is not
surprising that $1/L^{d-1}$ is the appropriate scaling variable in our
case. Close to the critical point, the distribution $\Pc$ scales with the
system size as~\cite{binder1981a, nicolaides1988a, bruce1992a}:
\begin{equation}
  P_L (\etac) = b_0 L^{\beta/\nu} {\cal P}^0 ( b_0 L^{\beta/\nu} \etac),
\end{equation}
where $P_L(\etac)$ is the distribution $\Pc$ measured in the finite system
of size $L$, $b_0$ some non--universal constant, and ${\cal P}^0$ a
function independent of system size. According to \eq{eq:fl}, the free
energy barrier $F_L$ is given by the peak--to--valley height in the
logarithm of $P_L(\etac)$. The above scaling property thus implies that
$F_L$ becomes $L$ independent close to the critical point. As a result,
the interfacial tension should scale with $1/L^{d-1}$.

\begin{figure}
\begin{center}
\includegraphics[clip=,width=\figwidth]{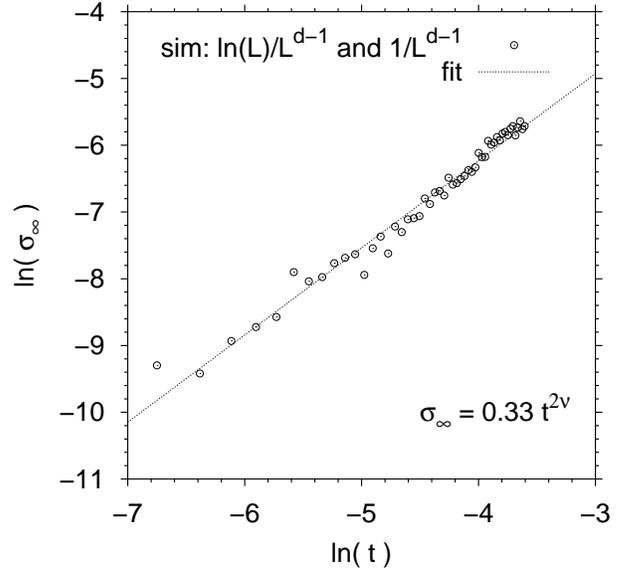}
\caption{~\label{fig16}The analogue of \fig{fig14} using two parameter
fits in $\ln(L)/L^{d-1}$ and $1/L^{d-1}$ when extrapolating the 
interfacial tension to the thermodynamic limit. The best collapse occurs 
\mbox{at $\etaprcr=0.7689$}. For the critical exponent we obtain 
$2\nu=1.30 \pm 0.04$ and for the critical amplitude $\sigma_0 = 0.33 \pm 
0.03$.}
\end{center}
\end{figure}

For completeness, we also show in \fig{fig16} the behavior of $\sigma_\infty$ as
function of $t$ when two--parameter fits in both $\ln(L)/L^{d-1}$ and
$1/L^{d-1}$ are used. Since only four different system sizes are considered, the
statistical quality of the extrapolations will likely deteriorate, and this
clearly shows in \fig{fig16}. Still, even then, it is possible to extract the
critical exponent and the critical point with reasonable accuracy. Note that the
deviations in $\etaprcr$ obtained using the various extrapolation schemes are
below the one percent level. In practice, the choice of extrapolation scheme
thus only becomes important when high resolution data are available.

\subsection{Critical amplitude ratios}

We now turn to the critical amplitude ratios that can be extracted from
\tab{tab:crit}. We first calculate $U_2$ given by \eq{eq:amp1}. If we
consider the colloidal phases we obtain $U_2 = 3.9 \pm 0.9$. The
corresponding ratio for the polymers is found to be $U_2 = 4.1 \pm 1.9$.  
Within the error bars, these estimates are compatible with \eq{eq:amp1}.  
The quantity $w^2 R_\sigma^{3/2}$ is found to be $0.10 \pm 0.04$ for the
colloidal phases, and $0.08 \pm 0.04$ for the polymer phases. This is
again compatible with \eq{eq:amp2}, although one must be aware of the
large error bars in our estimates. Finally, the quantity
$(R_\sigma^+)^{3/2} / Q_c$ is found to be $0.40 \pm 0.16$ for the
colloids, and $0.35 \pm 0.07$ for the polymers. In this case, we
systematically underestimate the values listed in \eq{eq:amp3}. Note,
however, that the discrepancy with the 3D Ising values is not too severe.  
Given the difficulty in general of measuring critical amplitudes, even in
the case of simple lattice models~\cite{pelissetto2002}, the agreement we
obtain is already quite remarkable.

\section{Summary and Outlook}

In summary, we have studied the critical behavior of the order parameter,
interfacial tension, susceptibility and coexistence diameter of the AO
model with colloid to polymer size ratio $q=0.8$. An important result is
that the critical exponent of the interfacial tension equals the expected
3D Ising value $2\nu \approx 1.26$. This critical behavior is consistent
with previous simulations~\cite{vink2004a, vink2004b}, and also with
experimental work~\cite{chen2000} in which 3D Ising critical behavior was
observed in real colloid--polymer mixtures. The critical behavior of the
order parameter and the susceptibility are also 3D Ising like. Our data
for the coexistence diameter is not conclusive. This is related to the
small value of the critical exponent $\alpha$, which makes it difficult to
accurately resolve the critical behavior from the simulation data. More
accurate simulations of larger systems are required to resolve this.

We found that the critical amplitude ratios obtained from our simulations
are compatible with the 3D Ising universality class. This confirms the
consistency of our data, and is encouraging considering the AO model is an
asymmetric binary mixture and therefore difficult to simulate. We
emphasize, however, that our estimates for the critical amplitudes should
be regarded as consistency checks only. Their accuracy cannot compete with
that obtained in, for example, direct simulations of the 3D Ising lattice
model.

We have demonstrated that finite size scaling methods are equally
applicable to rather complex systems like the AO model, and can be used to
obtain results with meaningful accuracy. Regarding finite size
extrapolations of the interfacial tension, our analysis is consistent with
\olcite{berg1993a}, in the sense that the most consistent fits are
obtained by ignoring the $\ln(L)$ dependent term in \eq{eq:fss}. For the
AO model, we also observed that by using the incorrect scaling variable,
the critical point in particular is not estimated correctly.

This work provides additional insight into the critical behavior of the AO
model. For a complete understanding, the critical behavior of the
correlation length should still be investigated. While the corresponding
critical exponent is of course known, namely $-\nu$, the critical
amplitude is not. This critical amplitude, however, cannot be extracted
from the probability distribution $\Pc$. The usual approach to study the
correlation length is to consider the static structure factor
instead~\cite{das2003a}. This requires additional simulations which are
beyond the scope of this work, and will be postponed to a future
publication.

\acknowledgments

We are grateful to the Deutsche Forschungsgemeinschaft for support
(TR6/A5) and to M. M\"{u}ller for many stimulating discussions. Generous
allocation of computer time on the JUMP cluster at the Forschungszentrum
J\"{u}lich GmbH is gracefully acknowledged. We also thank K. Dawson, R.
Evans, and M. Schmidt for stimulating remarks.

\bibliographystyle{prsty}

\end{document}